\documentstyle[prd,aps,floats]{revtex}
\begin{document}
\draft
\preprint{SUSSEX-AST 98/4-3, astro-ph/9804177}

%
%
\twocolumn[\hsize\textwidth\columnwidth\hsize\csname 
@twocolumnfalse\endcsname

\title{Assisted inflation}
\author{Andrew R.~Liddle, Anupam Mazumdar and Franz E.~Schunck}
\address{Astronomy Centre, University of Sussex, Falmer, Brighton BN1
9QJ,~~~U.~K.}  
\date{\today} 
\maketitle
\begin{abstract}
In inflationary scenarios with more than one scalar field, inflation may 
proceed even if each of the individual fields has a potential too steep for 
that field to sustain inflation on its own. We show that scalar fields with 
exponential potentials evolve so as to act cooperatively to assist inflation,
by finding solutions in which the energy densities of the different scalar 
fields evolve in fixed proportion. Such scaling solutions exist for an 
arbitrary number of scalar fields, with different slopes for the exponential 
potentials, and we show that these solutions are the unique late-time 
attractors for the evolution. We determine the density perturbation spectrum 
produced by such a period of 
inflation, and show that with multiple scalar fields the spectrum is closer 
to the scale-invariant than the spectrum that any of the fields would 
generate individually.
\end{abstract}

\pacs{PACS numbers: 98.80.Cq \hspace*{2.1cm} Sussex preprint SUSSEX-AST 
98/4-3, astro-ph/9804177}

\vskip2pc]


\section{Introduction}

The idea of cosmological inflation \cite{inf,LL93} is an attractive one, 
solving a range of otherwise troubling problems. Inflation is normally 
achieved by a period of the Universe's evolution during which the energy 
density is dominated by the potential energy of a scalar field. Although 
quite probably the early Universe contained several scalar fields, it is 
normally assumed that only one of these fields remained dynamically 
significant for a long time, with the others rapidly finding their way into 
the minima of their respective potential energies. 

In this paper we consider scalar fields with exponential potentials. These 
are already known to have interesting properties; for example, if one has a 
universe containing a perfect fluid and such a scalar field, then for a wide 
range of parameters the scalar field `mimics' the perfect fluid, adopting its 
equation of state \cite{Wetterich,WCL}. These scaling solutions are 
attractors \cite{CLW} at late times. The behaviour of such a field during an 
inflationary epoch has also been considered \cite{CLW}.

What was not considered in Ref.~\cite{CLW} is the effect of introducing a 
scalar field with an exponential potential on the other scalar field. The 
simplest example would be if the other field also possessed an exponential 
potential. Then the behaviour of both fields will be modified, since they 
feel only their own potential gradient, but experience, via the expansion, 
the frictional effect of all scalar fields present.

\section{Dynamics}
\label{s:2}

For simplicity, we begin by considering $m$ scalar fields, $\phi_i$, which 
each have an identical potential
\begin{equation}
V(\phi_i) = V_0 \exp \left( - \sqrt{\frac{16 \pi}{p}} \,
	\frac{\phi_i}{m_{{\rm Pl}}} \right) \,,
\end{equation}
where $m_{{\rm Pl}}$ is the Planck mass. Note that there is no direct 
coupling of the fields, which influence each other only via their effect on 
the expansion. The equations of motion are
\begin{eqnarray}
\label{Motion}
H^2 & = & \frac{8\pi}{3m_{{\rm Pl}}^2} \sum_{i=1}^m \left[ V(\phi_i) +
	\frac{1}{2} \dot{\phi}_i^2 \right] \,; \\
\ddot{\phi}_i & = & - 3 H \dot{\phi}_i - \frac{dV(\phi_i)}{d\phi_i} \,.
\end{eqnarray}
Our fields are combined additively; this is different from {\em 
soft inflation} \cite{soft}, where an exponential potential multiplies the 
potential of another scalar field.

If there is only a single scalar field, this leads to the well-known 
power-law solution \cite{LM85}
\begin{equation}
a(t) \propto t^p \,.
\end{equation}
This is inflationary only if $p > 1$, i.e.~for sufficiently shallow 
exponentials. The power-law solution also applies for any $p$ in the range 
$1/3$ to $1$, where it is non-inflationary. For $p < 1/3$, the asymptotic 
solution is that of a free scalar field, with $a \propto t^{1/3}$ regardless 
of the value of $p$ in the range $(0,1/3)$.

We first find a particular solution where all the scalar fields are equal: 
$\phi_1 = \phi_2 = \cdots = \phi_m$. We shall later show it is the unique 
late-time attractor. With this ansatz the equations become
\begin{eqnarray}
H^2 & = & \frac{8\pi}{3m_{{\rm Pl}}^2} \, m \, \left[ V(\phi_1) +
	\frac{1}{2} \dot{\phi}_1^2 \right] \,; \\
\ddot{\phi}_1 & = & - 3 H \dot{\phi}_1 - \frac{dV(\phi_1)}{d\phi_1} \,.
\end{eqnarray}
These can be mapped to the equations of a model with a single scalar field 
$\tilde{\phi}$ by the redefinitions 
\begin{equation}
\label{scal}
\tilde{\phi}_1^2 = m \, \phi_1^2 \quad ; \quad  \tilde{V} = m\, V
	\quad ; \quad \tilde{p}	= mp \,,
\end{equation}
so the expansion rate is $a \propto t^{\tilde{p}}$, provided that $\tilde{p} 
> 1/3$. The expansion 
becomes quicker the more scalar fields there are. And in particular, 
potentials with $p < 1$, which for a single field are unable to support 
inflation, can do so as long as there are enough scalar fields to make 
$mp>1$. Note also that this solution does not require $p$ to exceed 1/3, only 
the product $mp$. If $mp$ is less than one third then the solution will 
instead be that of a free scalar field. 

Although the solution with all scalar fields equal is a particular one, it is 
in fact the generic late-time attractor. To see this, keep $\phi_1$ but 
replace the rest with the redefined fields 
\begin{equation}
\psi_i = \phi_i - \phi_1 \,, \quad i=2,\dots,m \,.
\end{equation}
These fields obey the equation
\begin{eqnarray}
\label{eq9}
\ddot{\psi_i} + 3H\dot{\psi}_i & = & \frac{V_0}{m_{{\rm Pl}}} 
	\sqrt{\frac{16\pi}{p}} \exp \left(- \sqrt{\frac{16\pi}{p}} \, 
	\frac{\phi_1}{m_{{\rm Pl}}} \right) \\ 
 & & \hspace*{-15pt} \times \left[ \exp \left( - \sqrt{\frac{16\pi}{p}} \, 
	\frac{\psi_i}{m_{{\rm Pl}}} \right) -1 \right]  \,, 
	\quad i = 2,\dots,m \,.\nonumber 
\end{eqnarray}
This is the equation of a scalar field in an effective potential
\begin{equation}
\ddot{\psi_i} + 3H\dot{\psi}_i = - \frac{\partial
	V_{{\rm eff}}(\phi_1,\psi_i)}{\partial \psi_i} \,,
\end{equation}
with
\begin{eqnarray}
V_{{\rm eff}} & = & V_0	\sqrt{\frac{16\pi}{p}} \exp \left(-
	\sqrt{\frac{16\pi}{p}} \, \frac{\phi_1}{m_{{\rm Pl}}} \right) 
	\nonumber \\
 & & \times \left[ \sqrt{\frac{p}{16\pi}} \, \exp \left( -
 	\sqrt{\frac{16\pi}{p}} \, \frac{\psi_i}{m_{{\rm Pl}}} \right) +
 	\frac{\psi_i}{m_{{\rm Pl}}}\right] \,.
\end{eqnarray}
The minimum in the $\psi_i$ direction is always at $\psi_i = 0$, regardless 
of the behaviour of $\phi_1$, so the late-time solution has all the $\phi_i$ 
equal. The length of time to reach this attractor will depend on the initial 
separation (the value of $\psi_i$) and the extent to which friction, coming 
from the expansion rate $H$, is important.

\section{Density perturbations}

It is now well known how to calculate the density perturbation produced in 
multi-scalar field models. Sasaki and Stewart \cite{SS} (see also 
\cite{YSGW}) quote the result
\begin{equation}
{\cal P}_{\cal R} = \left( \frac{H}{2\pi} \right)^2 \,
	\frac{\partial N}{\partial \phi_i} \, 
	\frac{\partial N}{\partial \phi_j}  \, \delta_{ij} \,,
\end{equation}
where ${\cal P}_{\cal R}$ is the spectrum of the curvature perturbation 
${\cal R}$ in the usual units \cite{LL93}, $N$ is the number of 
$e$-foldings of inflationary expansion remaining, and there is a summation 
over $i$ and $j$. Since $N = - \int H \, dt$, we have
\begin{equation}
\sum_i \frac{\partial N}{\partial \phi_i} \dot{\phi}_i = -H \,,
\end{equation}
where in our case each term in the sum is the same, yielding 
\begin{equation}
{\cal P}_{\cal R} = \left( \frac{H}{2\pi} \right)^2 \, \frac{1}{m} \,
	\frac{H^2}{\dot{\phi}_1^2} \,.
\end{equation}
Note that this last expression only contains one of the scalar fields, chosen 
arbitrarily to be $\phi_1$. This expression looks as if it is $m$ times 
{\em smaller} than the usual formula for a single scalar field (see 
e.g.~Ref.~\cite{LL93}); however, remember that the presence of multiple 
fields has modified both $H$ and $\dot{\phi}_1$.

Of particular interest is the spectral index $n$. This is given by \cite{SS}
\begin{equation}
\label{Spectral}
n -1 = 2\frac{\dot{H}}{H^{2}} - 2\frac{\frac{\partial N}{\partial \phi_i}
	\left(\frac{8\pi}{m^{2}_{\rm Pl}} \frac{\dot{\phi_i}
	\dot{\phi_j}}{H^{2}} \, - \frac{m^{2}_{\rm Pl}}{8\pi}
	\frac{V_{,i,j}}{V} \right) \, \frac{\partial N}
	{\partial \phi_j} }{ \delta_{ij} \frac{\partial N}{\partial 
	\phi_i} \, \frac{\partial N}{\partial \phi_j}} \,,
\end{equation}
where there is a summation over repeated indices and the commas indicate
derivatives with respect to the corresponding field component.
Under our assumptions, the complicated second term on the right-hand side of 
the above equation cancels out, and 
Eq.~(\ref{Spectral}) reduces to the simple form 
\begin{equation}
\label{Spec}
1-n = - 2\frac{\dot{H}}{H^{2}} = \frac{m^{2}_{\rm Pl}}{8\pi} \,
	\left[\frac{\frac{\partial V(\phi_1)}{\partial \phi_1}}{V(\phi_1)}
	\right]^2 = \frac{2}{m p} \,.
\end{equation}

This result shows that the spectral index also matches that produced by a 
single scalar field with $\tilde{p} = mp$. The more scalar fields there are, 
the closer to scale-invariance is the spectrum that they produce. Note 
however that if the fields have such steep potentials as to be individually 
non-inflationary, $p < 1$, then many fields are needed before the spectrum is 
flat enough (say $n>0.7$) to have the possibility of explaining the observed 
structures. Large numbers of scalar fields are predicted by some theories, 
for example the 70 scalar fields, with unknown potentials, of the low-energy 
compactified superstring effective action \cite{CJ}.


\section{Potentials with different slopes}

We now generalize the above discussion, by considering each potential to have 
a different slope $p_i$
\begin{equation}
V_i(\phi_i) = V_0 \exp \left( - \sqrt{\frac{16 \pi}{p_i}} \,
        \frac{\phi_i}{m_{{\rm Pl}}} \right) .
\end{equation}
Notice that we keep the same $V_0$ for each field; since changing $V_0$ is 
equivalent to shifting the scalar field definition by a constant, this serves 
to fix the zero values of the fields.

We conjecture that scaling solutions exist, where the energy densities of the 
different fields attain fixed ratios at late times, relative to an 
arbitrarily chosen field $\phi_1$: 
\begin{equation}
\label{dotphi}
\frac{\dot\phi_i^2}{\dot\phi_1^2} =\frac{V_i(\phi_i)}{V_1(\phi_1)} = C_i \,.
\end{equation}
To guess the appropriate form of $C_i$, we note that the slow-roll 
approximation 
gives
\begin{equation}
\dot\phi_i^{2} \simeq \frac{2}{3 p_i}\frac{ V_i^2 (\phi_i) }{\sum_i
	V_i(\phi_i)} \,,
\end{equation}
which suggests 
\begin{equation}
C_i = \frac{p_i}{p_1} \,.
\end{equation}
We stress though that the slow-roll approximation is not needed in what 
follows.

Integrating the kinetic part of Eq.~(\ref{dotphi}) then gives
\begin{equation}
\label{phirat}
\phi_i = \sqrt{\frac{p_i}{p_1}} \, \phi_1 + \alpha_i \,,
\end{equation}
where $\alpha_i$ are the integration constants.
Ensuring the potentials also scale as in Eq.~(\ref{dotphi}) requires the 
constants to have values
\begin{equation}
\alpha_i = - \sqrt{\frac{p_i}{16\pi}} \, m_{{\rm Pl}} \, \ln
	\frac{p_i}{p_1} \,.
\end{equation}

To see that this solution will solve the full dynamical equations, we 
generalize 
the scaling argument of Section \ref{s:2}. 
Eq.~(\ref{phirat}) reduces us to a single degree of freedom $\phi_1$ in a 
manner consistent with the equations of motion, which become
\begin{eqnarray}
H^2 & = & \frac{8\pi}{3m_{{\rm Pl}}^2} \, \frac{\sum_{i=1}^m p_i}{p_1} \,
	\left[ V_1(\phi_1) + \frac{1}{2} \dot{\phi}_1^2 \right] \,; \\
\ddot{\phi}_1 & = & - 3 H \dot{\phi}_1 - \frac{dV_1(\phi_1)}{d\phi_1} \,.
\end{eqnarray}
Using the scaling of the potential from Eq.~(\ref{dotphi}), the other scalar 
wave equations all match the latter of these, confirming consistency of the 
ansatz. Note in particular that Eq.~(\ref{phirat}) brings all the 
exponentials into the same form.

These can then be turned into the equations of a model with a single scalar 
field via the redefinitions
\begin{equation}
\tilde{\phi}_1^2 = \frac{\sum p_i}{p_1} \, \phi_1^2 \quad ; \quad 
	\tilde{V_1} = \frac{\sum p_i}{p_1} \, V_1 \quad ; \quad \tilde{p}
	= \sum_i p_i \,,
\end{equation}
of which Eq.~(\ref{scal}) is a special case. This result is exact, not 
requiring a slow-roll approximation, and once more shows that the presence of 
multiple scalar fields increases the expansion rate. The expansion law is $a 
\propto t^{\tilde{p}}$, and is valid provided that $\tilde{p} > 1/3$.

The scaling construction shows that this solution exactly solves the 
multi-scalar field equations. One can show that this solution is an attractor 
by 
generalizing the argument of Section~II, via the ansatz 
\begin{equation}
\psi_i = \phi_i - \sqrt{\frac{p_i}{p_1}} \, \phi_1 - \alpha_i \,, 
	\quad i=2,\dots,m \,.
\end{equation}
which generalizes Eq.~(\ref{eq9}) to 
\begin{eqnarray}
\ddot{\psi_i} + 3H\dot{\psi}_i & = & \frac{V_0}{m_{{\rm Pl}}} 
        \sqrt{\frac{16\pi}{p_i}} \exp \left(- \sqrt{\frac{16\pi}{p_1}} \, 
        \frac{\phi_1}{m_{{\rm Pl}}} \right) \\ 
 & & \hspace*{-27pt} \times \frac{p_i}{p_1} \left[ \exp \left( -
 	\sqrt{\frac{16\pi}{p_i}} \, \frac{\psi_i}{m_{{\rm Pl}}} \right)-1 
 	\right]  \,, \quad i = 2,\dots,m \,. \nonumber 
\end{eqnarray}
As before, the effective potentials for the $\psi_i$ fields have a unique 
minimum at $\psi_i = 0$ for all $i=2,\dots,m$. Our scaling solution is 
therefore 
the unique late-time attractor.\footnote{We have also confirmed these 
solutions 
as late-time attractors numerically for a wide range of values of $p_i$.} 

The calculation of the spectral index follows the same lines as before, 
yielding 
\begin{equation}
1-n = \frac{2}{\tilde{p}} \,.
\end{equation}
This reduces to Eq.~(\ref{Spec}) when the slopes of the potentials are same.

\section{Conclusion}

Although the early Universe is likely to contain many scalar fields, a common 
assumption when analyzing inflation is that all but one of these fields has 
become dynamically irrelevant. However, for scalar fields with exponential 
potentials the late-time behaviour is for the energy densities of the 
different fields to scale with each other, as had already been noted for the 
case of a scalar field with an exponential potential plus a barotropic fluid 
\cite{Wetterich,WCL,CLW}, even if the fields have no direct coupling to each 
other and if their potentials have different slopes.

Such multiple scalar fields can act cooperatively to drive a period of 
inflation, even if the individual fields have potentials which are too steep 
in their own right; the expansion law in the scaling solution is 
$t^{\tilde{p}}$, where $\tilde{p} = \sum p_i$ with the $p_i$ being the 
power-law expansion rates that the individual fields would drive in 
isolation. The reason for this behaviour is that while each field experiences 
the `downhill' force from its own potential, it feels the friction from all 
the scalar fields via their contribution to the expansion rate.

We have also studied the density perturbation spectrum produced, which has a 
spectral index $n$ matching that of power-law inflation driven by a single 
field at rate $\tilde{p}$. The spectrum is therefore brought closer to 
scale-invariance the more fields participate in the inflationary expansion. A 
perturbation spectrum close to scale invariance is preferred by current 
observations, and this phenomena may offer assistance to supergravity-based 
inflation models which often predict spectra which are not all that close to 
scale invariance \cite{Lyth}.


\section*{Acknowledgments}

A.R.L. is supported by the Royal Society, A.M. by the Inlaks foundation and
the ORS, and F.E.S. by the European Union TMR Marie Curie programme. We 
thank Ed Copeland, Jim Lidsey and David Wands for discussions, and 
acknowledge use of the Starlink computer system at the University of Sussex. 

\end{document}